\documentstyle[emulateapj]{article}

\newcommand{\hii}{H\,{\sc ii}}

\newcommand{\siot}{SiO $(2 \rightarrow 1)$}

\newcommand{\siof}{SiO $(5 \rightarrow 4)$}

\newcommand{\methanol}{CH$_3$OH}

\newcommand{\thcotw}{$^{13}$CO ${(2 \rightarrow 1)}$}
\newcommand{\hctn}{HC$_3$N ${(15 \rightarrow 14)}$}
\newcommand{\htwoo}{H$_2$O}
\newcommand{\htwo}{H$_2$ ($\nu = 1 \rightarrow 0$) S(1)}
\newcommand{\htw}{H$_2$}

\newcommand{\msun}{M$_{\sun}$}
\newcommand{\lsun}{L$_{\sun}$}

\newcommand{\um}{$\mu$m}
\newcommand{\cm}{cm$^{-3}$}
\newcommand{\kms}{km~s$^{-1}$}

\newcommand{\brg}{Br-$\gamma$}
\begin{document}
\received{ }
\accepted{  }
\journalid{ }{ }
\slugcomment{   }
\lefthead{Megeath \& Tieftrunk}
\righthead{The Detection of Outflows}

\title{The Detection of Outflows in the IR-Quiet Molecular Core NGC
6334 I(North)\footnote{based on observations conducted at the
European Southern Observatory, La Silla, Chile}}

\author{S. T. Megeath}
\affil{Harvard--Smithsonian Center for Astrophysics, 60 Garden St.,
Cambridge, MA 02138}
\author{A. R. Tieftrunk}
\affil{K\"olner Observatorium f\"ur Submm--Astronomie, 
1. Physikalisches Institut der Universit\"at zu K\"oln,
Z\"uplicher Str. 77, D-50937 K\"oln, Germany 
and ESO/La Silla, Casilla 19001, Santiago 19, Chile}



\begin{abstract}

We find strong evidence for outflows originating in the dense
molecular core NGC 6334 I(North): a $1000$~\msun\ molecular core
distinguished by its lack of \hii\ regions and mid--IR emission.  New
observations were obtained of the \siot\ and $(5 \rightarrow 4)$ lines
with the SEST 15--m telescope and the \htwo\ line with the ESO 2.2--m
telescope.  The line profiles of the SiO transitions show broad wings
extending from $-50$ to $40$~\kms, and spatial maps of the line wing
emission exhibit a bipolar morphology with the peaks of the red and
blue wing separated by $30''$.  The estimated mass loss rate of the
outflow is comparable to those for young intermediate to high--mass
stars.  The near--IR images show eight knots of \htw\ emission.
Five of the knots form a linear chain which is displaced from the
axis of the SiO outflow; these knots may trace shock excited gas along
the path of a second, highly collimated outflow.  We propose that I(N)
is a rare example of a molecular core in an early stage of cluster
formation.

\end{abstract}


\section{Introduction}

At a distance of 1.7~Kpc (Neckel 1978), the NGC 6334 molecular cloud
is an 11~pc long filament containing a remarkable chain of five
luminous star--forming regions (McBreen et al.~1979).  The youngest of
these star--forming regions is thought to be the northernmost of the
chain, designated NGC 6334 I in the McBreen et al. nomenclature.
Evidence for the youth of this region is found in the sheer
concentration of phenomena thought to accompany the earliest stages of
star formation: an array of OH, \htwoo, \methanol\ and NH$_3$(3,3)
masers (Gaume \& Mutel 1987; Forster \& Caswell 1989; Menten \& Batrla
1989; Kraemer \& Jackson 1995), luminous mid--IR sources (Harvey \&
Gately 1983; Kraemer et al.~1999), an ultracompact \hii\ region
(DePree~et~al.~1995), a bipolar outflow (Bachiller \& Cernicharo 1990;
Davis \& Eisloeffel 1995; Persi et al. 1996), and a cluster of stars
detected at 1-2~\um\ (Tapia, Persi \& Roth 1996).

The young stars of NGC 6334 I are embedded in a massive dense
molecular core which has been mapped in a number of spectral lines and
the submillimeter continuum.  Interestingly, these same maps show that
NGC 6334 I is part of a twin core system: only $2'$ north (1 pc at 1.7
kpc) of NGC 6334 I, a second dense core of roughly equal size is
apparent.  This second core, designated NGC 6334 I(North) [hereafter:
I(N)] by Gezari (1982), has perhaps the strongest NH$_3$ lines
observed in the sky, yielding a gas temperature of 30~K and a volume
density in excess of 10$^6$~\cm\ (Forster~et~al. 1987; Kuiper et
al. 1995). Estimates of the total mass of the cloud range from 1000 to
3000~\msun (Gezari 1982; Kuiper et al. 1995).  However, in distinct
contrast to source I, NGC 6334 I(N) shows no detectable H\,{\sc ii}
regions, mid--IR emission, or clusters of 2~\um\ sources
(Loughran~et~al. 1986; Ellingsen, Norris \& McCulloch 1996; Tapia,
Persi \& Roth 1996).

Radio observations have detected an H$_2$O maser, a cluster of Class I
\methanol\ masers, and a Class II \methanol\ maser toward I(N) (Moran
\& Rodriguez 1980; Kogan \& Slysh 1998; Walsh~et~al. 1998).  The
enigmatic presence of masers without associated infrared emission or
H\,{\sc ii} regions has lead to the conclusions that I(N) is a
pre--stellar core about to undergo star formation (Kuiper et al. 1995)
or that I(N) contains a protostar in a pure accretion phase (Moran \&
Rodriguez 1980).  In this letter, however, we present strong evidence
for outflows originating in I(N), indicating that the I(N) core
already contains at least one protostar capable of driving a powerful
outflow.

\section{Observations and Data Reduction}

The near--IR data were obtained over three nights in June 1998
with the IRAC2b camera on the ESO 2.2--m telescope on La~Silla, Chile.
The weather was photometric and the seeing was $1''$. The detector was
a $256 \times 256$ pixel NICMOS3 array with a pixel scale of
0.5\arcsec. A warm Fabry--Perot (FP) interferometer with a spectral
resolution of $\lambda / \delta \lambda \sim 1000$ was used with
narrow band filters acting as order sorters.  Toward 16 on--source and
reference positions, we obtained 3 minute integrations at ten
wavelengths, including three wavelengths bracketing the \htwo\ line,
three bracketing the \brg\ line, and four wavelengths selected to
measure the continuum contribution.  Calibration standards were
imaged at each of the FP settings.  To avoid internal reflections from
the FP, dome flat fields were obtained for each narrow band filter
with the FP out of the optical path. The data were reduced with a
custom program written in the IDL environment.

The millimeter--wave observations were carried out in May and June
1998 using the 15--m Swedish-ESO Submillimeter Telescope (SEST). The
observations were made using SIS receivers with $T_{\rm sys}$ of
100-200 K and 1440 channel acousto--optical spectrometers. The \siot\
and $(5 \rightarrow 4)$ lines were mapped simultaneously using the
dual beam--switching mode with a beam throw of $12'$ in azimuth.  The
\thcotw\ line was subsequently mapped in position switching mode with
a reference position offset by $(\Delta \alpha, \Delta \delta) =
(-20',20')$. In September 1999, $^{12}$CO ($2 \rightarrow 1$) spectra
were obtained toward three positions; position switching was used with
an offset of $(\Delta \alpha, \Delta \delta) = (30',0')$.  The FWHP
beamwidths are $58''$ for \siot\ and $23''$ for \siof, \thcotw\ and
$^{12}$CO ($2 \rightarrow 1$).  The velocity widths per channel
(typical RMS noises per channel) are 2.39 (0.03), 0.97 (0.04), 0.95
(0.10) and 0.95~\kms\ (0.12~K) for the \siot, \siof, \thcotw, and
$^{12}$CO ($2 \rightarrow 1$) lines, respectively.  Using the CLASS
package, 1st and 3rd order baselines were subtracted from the
SiO and CO data, respectively. The temperatures were calibrated using
the chopper--wheel method and placed on a main--beam brightness
temperature scale using the beam efficiences given in the SEST
Handbook ver2.1 (1998).  The pointing corrections never exceeded
$5''$.

\section{Analysis}

We mapped the \siof\ and $(2 \rightarrow 1)$ lines in a $5 \times 5$
grid with 20\arcsec\ spacings centered on the I(N) core.  SiO line
emission is detected in both transitions at all 25 map positions.
Broad wings are detected in both the $(2 \rightarrow 1)$ and $(5
\rightarrow 4)$ spectra at velocities ranging from $V_{lsr} =
-50$~\kms\ to 40~\kms.  In Fig.~\ref{fig:spectra}, we display \siot\
and ($5 \rightarrow 4$) spectra toward the positions showing the
strongest blue and red wing emission.

\begin{figure*}[t]
\plotone{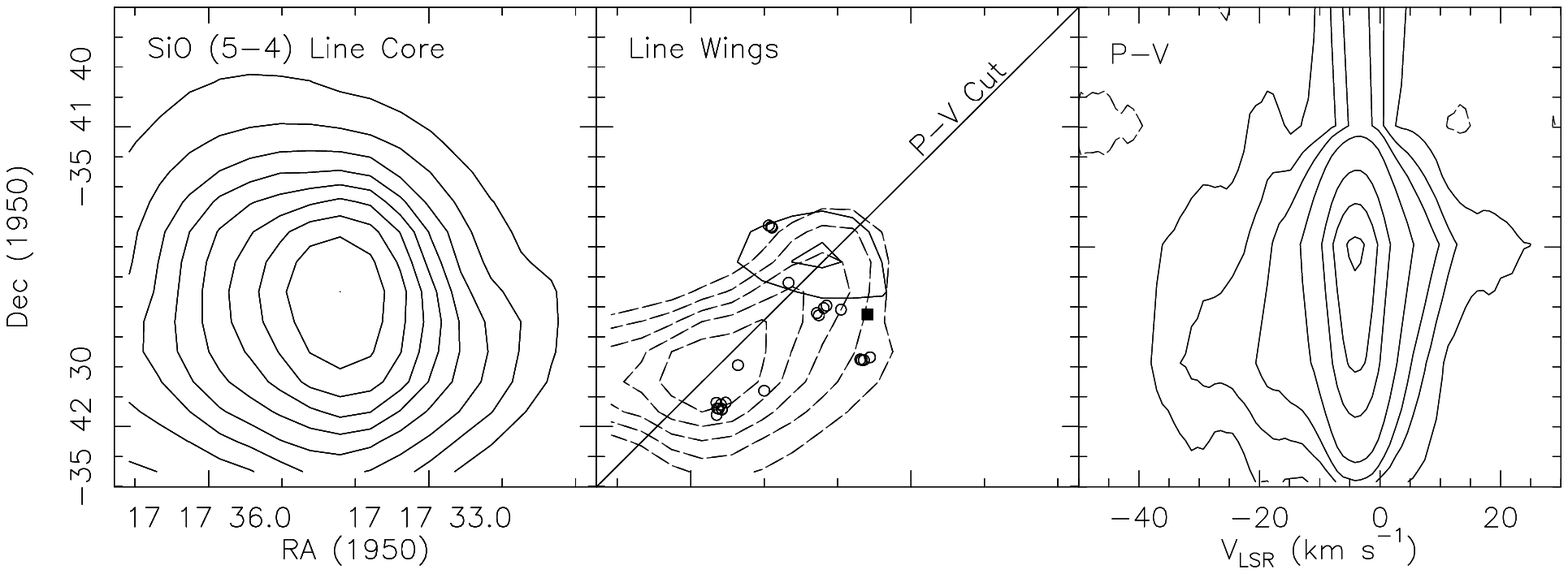}
\caption[] {The \siof\ line emission. Left panel: the line core
emission integrated over $V_{lsr} = -15$ to 7.5~\kms. The contour
levels are 3.9 to 17.55 by 1.95 K~\kms.  Middle panel: the dashed
contours show the blue--shifted line wing emission integrated over
$V_{lsr} = -50$ to $-15$~\kms\ and the solid contours show the
red--shifted gas integrated over $V_{lsr} = 7.5$ to $40$~\kms. The
contour levels are 2.5 to 4.5 by 0.5~K~\kms.  The open circles mark
the 44 GHz Class I methanol masers (Kogan \& Slysh 1998), and the
filled square marks the 6 and 12 GHz Class II methanol maser
(Walsh~et~al. 1998).  Right panel: the position--velocity diagram
along the the diagonal cut shown in the middle panel.  The velocity
resolution is smoothed to 4.9~\kms.  The contour levels are -0.03,
0.07, 0.14, 0.21, 0.35 to 1.40 by 0.35~K}
\label{fig:contour} 
\end{figure*}

The velocity integrated \siof\ emission of the red wing, blue wing,
and line core are plotted in Fig~\ref{fig:contour}.  The wings are
integrated over the velocity range in which wing emission is apparent
in {\it both} the \siof\ and $(2 \rightarrow 1)$ lines and is clearly
separable from the line core.  Higher spectral resolution \siot\ and
$(3 \rightarrow 2)$ data show the wings extending into the -15 to
7.5~\kms\ velocity interval of the plotted line core emission
(Tieftrunk \& Megeath 1999); however, this low velocity wing component
cannot be isolated in the low spectral resolution data presented here.

The SiO line core emission is sharply peaked at a position of
$\alpha_{1950} = 17^h 17^m 34.3^s$, $\delta_{1950} = -35^{\circ} 42'
10''$, coincident to within $10''$ of the peaks of the dense gas
tracers \hctn, CS ($7 \rightarrow 6$) and NH$_3$(3,3) (Megeath \&
Sollins 1999; Kraemer \& Jackson 1999) and of the peak of the 350 --
1100~\um\ continuum emission (Sandell 1999). The red wing peaks $10''$
to the north of the center of I(N) (as defined by the peak core
emission); the stronger, more spatially extended blue wing peaks
$20''$ to the southeast of the center.  The spatial offset of the red
and blue wings is also apparent in the position--velocity diagram
(Fig.~\ref{fig:contour}). The positions of the peaks are not dependent
on the adopted integration intervals or the choice of contours.  The
detection of broad wings which peak within $20''$ of the center of
I(N), and which exhibit a spatial offset between the blue and
red--shifted emission peaks, is strong evidence for a bipolar outflow
originating in I(N).

\centerline{\includegraphics[]{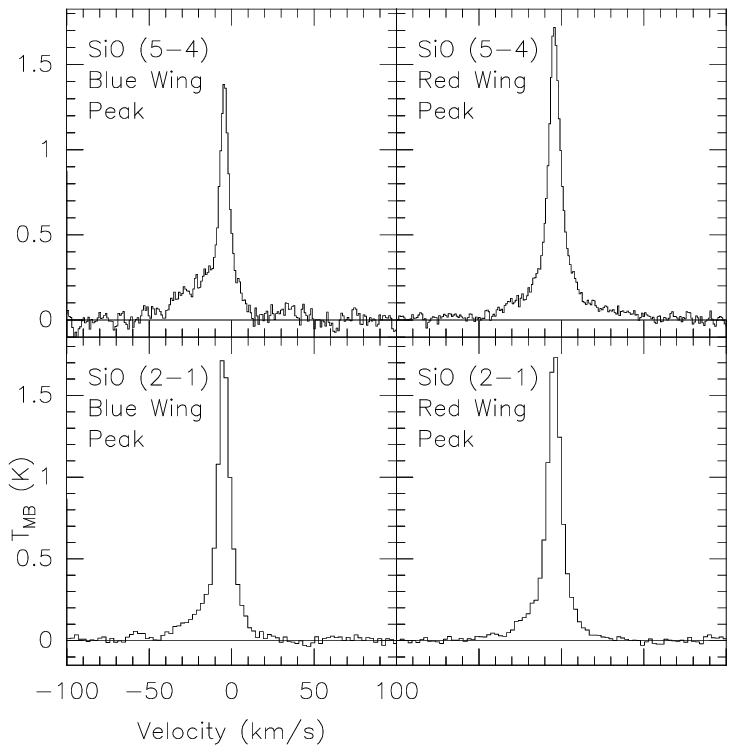}}
\vskip -2.3 in
\figcaption[]{Spectra of the \siot\ and \siof\ emission lines toward
NGC 6334 I(N). Displayed are the spectra at
the peak positions of the blue and red wings.\label{fig:spectra}}
\vskip 0.2 in

The column density of SiO in the outflow is determined independently
for each 5~\kms\ velocity interval from $V_{lsr} = -50$ to $-15$~\kms\
for the blue wing and from 10 to 40~\kms\ for the red wing.  Assuming
optically thin emission and convolving the \siof\ data to the beamsize
of the \siot\ data ($58''$), we calculate the column density in the
$J=2$ and $J=5$ states and estimate a rotational temperature from the
ratio of these column densities (Goldsmith \& Langer 1999).  The
rotational temperatures are $\sim 10~K$ for all of the velocity
intervals.  Assuming a constant rotational temperature between all of
the rotational states, a total SiO column density is determined for
each velocity interval.  The 10~K rotational temperature is
significantly less than the 30~K kinetic temperature of the core
(Kuiper et al. 1995), indicating that the SiO is subthermally excited.
A statistical equilibrium simulation with a large velocity gradient
approximation shows that for kinetic temperatures of 30 to 100~K and
volume densities of $10^6$~cm$^{-3}$ to $3 \times 10^5$~cm$^{-3}$, the
distribution of molecules among the rotational states can be well
approximated by a contant rotational temperature with a value of $\sim
10$~K.  With these simulations, we estimate that the errors incurred
by our assumption of a constant rotational temperature are less than
20\%.

The derivation of the outflow properties require an in situ
measurement of the abundance of SiO relative to H$_2$, as the SiO
abundance can be enhanced by orders of magnitude in outflows (Acord,
Walmsley \& Churchwell 1997; Codella, Bachiller \& Reipurth 1999).  We
use the \thcotw\ spectra to determine the column density of $H_2$.
Toward the peak position of the blue SiO line wing emission, the
\thcotw\ spectra shows a clear detection of the blue line wing. We
limit our analysis to the velocity range of $V_{lsr} = -20$ to
$-15$~\kms, over which the \thcotw\ wing has a high signal--to--noise
and shows no troughs indicative of contaminating emission in the
reference beam.  The ratio of the $^{12}$CO to $^{13}$CO temperatures
at this position and velocity range indicate that the $^{13}$CO is
optically thin with an optical depth of 0.1.  We adopt a kinetic
temperature of 30 K (Kuiper et al.  1995) and assume LTE, resulting in
a $^{13}$CO column density of $4.6 \times 10^{14}$~cm$^{-2}$ averaged
over a $58''$ FWHP beamwidth and integrated over the $-20$ to
$-15$~\kms\ velocity interval.  Assuming a $^{12}$C/$^{13}$C ratio of
59 (Wilson \& Rood 1994) and a [CO]/[H$_2$] abundance of $10^{-4}$,
the column density of $H_2$ is $2.7 \times 10^{20}$~cm$^{-2}$.  For
the identical velocity interval and beamwidth, the SiO column density
is $2.2 \times 10^{12}$~cm$^{-2}$ and the resulting relative abundance
is [SiO]/[H$_2$] = $8 \times 10^{-9}$.  This value is similar to that
derived in the outflow of the neighboring NGC 6334 I core by Bachiller
\& Cernicharo (1990).  The estimated abundance decreases with an
increase in the adopted kinetic temperature: a temperature of 100 K
yields [SiO]/[H$_2$] = $3 \times 10^{-9}$.  A similar analysis of the
SiO column density in the Gaussian--shaped line cores and a H$_2$
column density derived from C$^{18}$O measurements (Megeath \& Sollins
1999), results in an SiO relative abundance of $2 \times 10^{-10}$ for
the low--velocity and quiescent gas in I(N), showing that the SiO
abundance is enhanced in the high velocity gas.

\tabcaption{}
\smallskip
\centerline{
\begin{tabular}{lll} \hline \hline
Property & 
Blue Wing & 
Red Wing \cr \hline
Mass& 4.9 ~\msun & 2.0 ~\msun  \cr
Momentum& 110~\msun~\kms & 46~\msun~\kms \cr
Energy& $3.0 \times 10^{46}$ ergs &$ 1.2 \times 10^{46}$ ergs \cr
Mass Loss\tablenotemark{a} & $1.0 \times 10^{-3}$~\msun~yr$^{-1}$ &$0.4 \times 10^{-3}$~\msun~yr$^{-1}$  \cr
Luminosity\tablenotemark{ab}& 51~\lsun & 20~\lsun \cr 
\hline
\end{tabular}}
\noindent
$^{a}$Adopting an age of 5000 years (see text).
\noindent
$^{b}$The luminosity is the mechanical luminosity given by the energy over the age.
\vskip 0.25 in

Adopting a SiO relative abundance of $8 \times 10^{-9}$, we estimate
a total momentum and energy using the equations given in Choi, Evans
\& Jaffe (1993). To calculate the dynamical timescale, we take the
outflow radius to be $15''$ (or 0.12 pc), half the distance between
the emission peaks of the line wings in Fig.~\ref{fig:contour}, and
divide by the mass--weighted average velocity to obtain a value of
5000 years.  The derived outflow properties are listed in Table 1.
These values are not corrected by an assumed inclination angle.

Independent evidence for an outflow in I(N) is shown in our continuum
subtracted image of the 2.12~\um\ \htwo\ emission (Fig.~\ref{fig:h2}).
Through a careful inspection of the images, we have identified eight
knots of H$_2$ emission toward I(N).  The brightest of the knots is
the nebula detected in broad band 2.2~\micron\ imaging by Tapia, Persi
\& Roth (1996).  The lack of any known UV sources in I(N) capable of
fluorescing H$_2$ molecules, the distribution of the H$_2$ emission in
widely separated, compact knots, and the absence of continuum emission
toward the H$_2$ knots suggests that the knots are shock--excited
nebulae in one or more outflows.  The knots are to the northwest of
the SiO outflow, and they possibly trace the continuation of the
redshifted lobe of the outflow. However, five of the knots fall in a
linear chain which is displaced from the axis of the bipolar outflow
detected in SiO. This geometry suggests that these five knots may
trace a second, highly collimated outflow in I(N).

\section{Discussion}

The detection of the bipolar molecular outflow is strong evidence for
ongoing star formation in NGC 6334 I(N).  The presence of outflows,
and the implication that the core is not starless, provides an
explanation for the enigmatic presence of CH$_3$OH and H$_2$O
masers in the core.  Indeed, the detected masers are coincident with
the observed high velocity gas (Fig.~\ref{fig:contour}).  Since the
driving source of the SiO outflow has not been directly detected, the
source properties must be inferred from the properties of the outflow,
with the constraint that the source luminosity must be less than the
total core luminosity of $10^4$~\lsun\ derived from
submillimeter observations (Gezari 1982).  Shepherd \& Churchwell
(1996) have found a relationship between source luminosity
vs. mass loss rate for luminosities ranging from 1 to $10^5$~\lsun.

\vskip -0.1 in
\centerline{\includegraphics[]{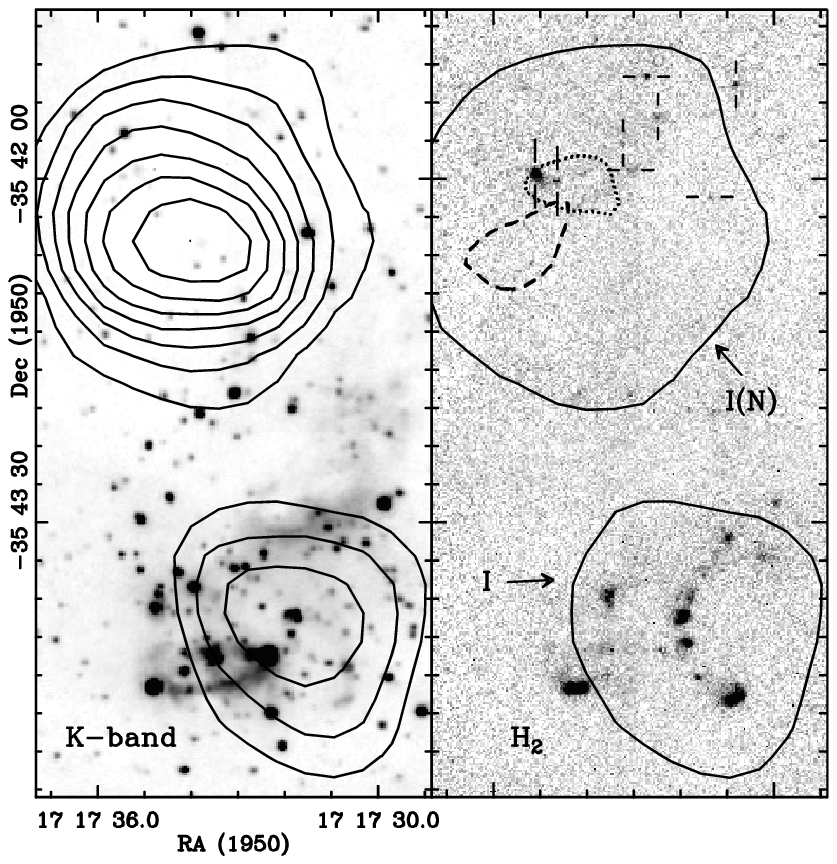}}

\vskip -0.5 in 

\figcaption[]{
Left panel: a $K$--band image with contours of \hctn\ emission
overlaid; the contour levels are 5.19 to 17.3 by 1.73 K~\kms (Megeath,
Sollins \& Wilson 1999).  The NGC 6334 I region is centered on the southern
HC$_3$N clump; the northern clump is the NGC 6334 I(N) core.  A dense
cluster of 2~\um\ sources is apparent toward I.  Right panel: the
continuum subtracted \htwo\ emission. Overlayed are the contours of
the blue (4.0 K~\kms; dashed) and red (2.5 K~\kms; dotted) \siof\ line
wings and the integrated \hctn\ emission (5.19 K~\kms; solid).  A
complex of H$_2$ emission knots is visible toward I.  Toward I(N), the
positions of the eight H$_2$ emission knots are marked.
The five knots tracing a possible collimated outflow are marked 
by the vertical lines.\label{fig:h2}}

\vskip 0.1 in

The mass loss rate of the NGC 6334 I(N) outflow,
$10^{-3}$~\msun~yr$^{-1}$, is characteristic of sources with
luminosities of $10^3$~\lsun.  Due to the unknown inclination of the
outflow, the actual mass loss rate can be substantially different;
however, it is unlikely that the resulting uncertainty in the
luminosity is more than a factor of ten. Consequently, the driving
source is probably a deeply embedded intermediate to high--mass
(proto)star.

It is likely that I(N) is forming multiple stars.  The geometry of the
H$_2$ knots is suggestive of at least one additional outflow, driven
by a second, embedded source.  The detection of five faint, highly
reddened ($H-K > 1.6$) near--IR point sources and three distinct
submillimeter condensations within the confines of the I(N) core is
further evidence for multiple sites of star formation (Tapia, Persi \&
Roth 1996; Sandell 1999).  Additionally, the mass, density and size of
the I(N) core are comparable to those of cluster--forming cores;
Fig.~\ref{fig:h2} shows that the similar NGC 6334 I core contains a
cluster of more than 90 stars (Tapia, Persi \& Roth 1996).  For these
reasons, we propose that NGC 6334 I(N) is a rare example of a
molecular core in the early stages of cluster formation.  We predict
that I(N) will evolve into a region similar to the active I core, with
a pronounced cluster of 2~\um\ sources, H\,{\sc ii} regions and
luminous mid--IR sources.  A longstanding question is whether star
formation in clusters progresses in a sequence, starting with the
low--mass stars and ending with the formation of high--mass stars. If
the NGC 6334 I(N) core is in the process of forming a cluster, then
the formation of at least one intermediate to high--mass star has
preceeded the {\it appearance} of a dense cluster of low--mass stars.
This does not necessarily imply that high--mass star formation has
preceeded the {\it formation} of most of the low--mass stars; the
low--mass stars may be hidden by the high extinction in the core.
These results do imply that the earliest stages of cluster formation
may be identified not by bright mid--IR emission, \hii\ regions or
dense clusters of 2~\um\ sources, but by the presence of powerful
outflows and masers detectable at radio wavelengths.  Surveys for
masers and outflows, particularly in regions without mid--IR sources,
is a promising means for identifying molecular cores at the onset of
cluster formation.

\acknowledgements

We thank S. H\"uttemeister and T. Bergin for providing results from
their LVG codes, P. Myers, P. Sollins and Q. Zhang for commenting on
this manuscript, and K. Kraemer, J.  Jackson, K. Menten, G. Sandell
and P. Pratap for valuable discussions.

\newpage
			     

\begin{thebibliography}{ }
\bibitem[]{a1} Acord, J., Walmsley, C. M. \& Churchwell, E. 1997, \apj,
475, 693

\bibitem[]{b1} Bachiller R. \& Cernicharo, P. 1990, A\&A, 239, 276




\bibitem[]{c3} Choi, M., Evans II, N. J. \& Jaffe, D. T. 1993, \apj, 417, 624

\bibitem[]{c4} Codella, C., Bachiller, R. \& Reipurth, B. 1999, A\&A, 343,
585

\bibitem[]{d1} Davis, C. J. \& Eisloeffel, J. 1995, \apj, 300, 851

\bibitem[]{d2} DePree, C. G., Rodriguez, L. F., Dickel, H. R. \&
Goss, W. M. 1995, \apj, 447, 220

\bibitem[]{d3} Ellingsen, S. P., Norris, R. P. \& McCulloch, P.M.
1996, MNRAS 279, 101

\bibitem[]{f1} Forster, J. R., Whiteoak, J. B., Gardner, F. F., 
Peters, W. L. \& Kuipers, T. B. H. 1987, Proc. ASA, 7, 189

\bibitem[]{f2} Forster, J. R. \& Caswell, J. L. 1989, A\&A, 213, 339

\bibitem[]{g1} Gaume, R. A. \& Mutel, R. B. 1987, \apjs, 65, 193

\bibitem[]{g2} Gezari, D. Y. 1982, 259, L29

\bibitem[]{g3} Goldsmith, P. F. \& Langer, W. D. 1999, \apj, 517, 209


\bibitem[]{h2} Harvey, P. M. \& Gately, I. 1983, \apj, 269, 613

\bibitem[]{k1} Kogan, L. \& Slysh, V. 1998, \apj, 497, 800

\bibitem[]{k2} Kraemer, K. E. \& Jackson, J. M. 1995, \apj, 439, L9

\bibitem[]{k2b} Kraemer, K. E., Deutsch, L. K., Jackson, J. M., Hora,
J. L., Fazio, G. G., Hoffmann, W. F. \& Dayal, A. 1999, \apj, 516, 817

\bibitem[]{k2c} Kraemer, K. E. \& Jackson, J. M. 1999, \apj, in press.

\bibitem[]{k3} Kuiper, T. B. H., Peters III, W. L., Foster, J. R.,
Gardner F. F. \& Whiteoak, J. B. 1995, \apj, 446, 692



\bibitem[]{l2} Loughran, L, McBreen, B., Fazio, G. G., Rengarajan,
T. N., Maxson, C. W., Serio, S., Sciortino, S. \& Ray, T. P. 1986, 
\apj, 303, 629

\bibitem[]{m1} McBreen, B., Fazio, G. G., Steir, M. \& Wright, E. L.
1979, \apj, 232, L183

\bibitem[]{m2} Megeath, S. T., Sollins, P., \& Wilson 1999, in prep.


\bibitem[]{m3} Menten, K. M. \& Batrla, W. 1989, \apj, 341, 839

\bibitem[]{m4} Moran, J. M. \& Rodriguez, L. F. 1980, \apj, 236, L159

\bibitem[]{n1} Neckel, T. 1978, A\&A, 69, 51


\bibitem[]{p1} Persi, P., Roth, M., Tapia, M., Marenzi, A. R., 
Felli, M., Testi, L. \& Ferrari-Toniolo, M. 1996, A\&A, 307, 591 


\bibitem[]{s1} Sandell, G. 1999, A\&A, in press.


\bibitem[]{s2} Shepherd, D. S. \& Churchwell, E. 1996, \apj, 472, 225

\bibitem[]{t1} Tapia, M., Persi, P. \& Roth, M. 1996, A\&A, 316, 102

\bibitem[]{t2} Tieftrunk, A. R. \& Megeath, S. T. 1999, in prep.



\bibitem[]{w0} Walsh, A. J., Burton, M. G., Hyland, A. R., \& Robinson,
G. 1998, MNRAS, 301, 640

\bibitem[]{w1} Wilson, T. L \& Rood, R. T. 1994, ARA\&A, 32, 192


\end{thebibliography}
\end{document}